\documentclass[10pt]{iopart}
\usepackage{graphicx}
\usepackage{siunitx}
\usepackage{iopams}
\begin{document}
\title[On the suitablility of rigorous coupled-wave analysis for fast optical force simulations]{On the suitability of rigorous coupled-wave analysis for fast optical force simulations}

\author{Bo Gao$^1$, Henkjan Gersen$^{1,2}$ and Simon Hanna$^1$}
\address{$^1$ H.H. Wills Physics Laboratory, University of Bristol, BS8~1TL, Bristol, UK}
\address{$2$ iLoF - Intelligent Lab on Fibre, Porto 4300-240, Portugal}
\ead{s.hanna@bristol.ac.uk}

\begin{abstract}
Optical force responses underpin nanophotonic actuator design, which requires a large number of force simulations to optimize structures.
Commonly used computation methods, such as the finite-difference time-domain (FDTD) method, are resource intensive and require large amounts of calculation time when multiple structures need to be compared during optimization.
This research demonstrates that performing optical force calculations on periodic structures using the rigorous coupled-wave analysis method is typically on the order of 10 times faster than FDTD with sufficient accuracy to suit optical design purposes.
Moreover, this speed increase is available on consumer grade laptops with a CUDA-compatible GPU avoiding the need for a high performance computing resource.
\end{abstract}

\noindent{\it Keywords}: optical force, rigorous coupled-wave analysis (RCWA), finite-difference time-domain (FDTD), metamaterials

\submitto{\JOPT}
\maketitle
\ioptwocol

\section{Introduction}

With the rapid development of nanofabrication techniques and increased understanding of optical responses, fabricating fast optically driven actuators becomes feasible. One of the challenges in practice is to design the structure of a potential actuator in a given material system with the desired optical response. This process is also known as inverse design\cite{InverseDesign}, and it is commonly carried out by comparing simulated responses for various structures.

A diverse range of conventional optimization algorithms have been used to accelerate the search for optimal structures in inverse design, including genetic algorithms\cite{GA}, simulated annealing\cite{SimulatedAnnealing}, gradient-search algorithms\cite{Gradient} and topology optimization\cite{Topo}. These optimization methods are search strategy driven, because the structures to be simulated depend on decisions made during the algorithm and cannot be foretold and prepared in advance. Recently, data-driven algorithms involving machine learning\cite{DeepLearning} and neural networks\cite{NeuralNetwork} have started to be used\cite{MLReview}, where the simulations needed for the training sets are independent of the optimization routines and can be prepared in advance and reused. 

Although the approaches mentioned above provide optimization strategies that avoid exploring the entire structural parameter space, they still require a large number of simulations of various structures. These simulations are normally the most computationally expensive parts in inverse design routines due to the high complexity of the structural parameters. As a result, the suitability of any inverse design method strongly depends on the speed and required computing resources to obtain an accurate simulation of an individual structure.

The interest in optical forces for actuation control has exploded in recent decades due to the development of optical trapping techniques\cite{Tweezers}. As a tool for manipulating the dynamic behaviour of nanophotonic structures\cite{LOF,LOF-probe,OT-probe}, optical forces have the potential for creating fast directly driven actuators\cite{AppReview}. Although the magnitude of optical forces is usually small, of the order of \unit{pN}, and operate on the microscopic scale, research shows that they can be manipulated and amplified to be observed at a macroscopic scale\cite{LOF-observe} typically through the use of repeats of a designed motif\cite{motif}. Although many kinds of optical forces are considered in designing actuators such as radiation pressure and gradient forces\cite{Gradient-Actuator}, we are particularly interested in optimising so-called lateral optical forces, which typically act in a direction that is not parallel to the propagation direction of the incident light. Recent examples of lateral optical forces include the design of plasmonic linear nanomotors\cite{LOF-nanomotor}, microscopic metavehicles\cite{benchmark}, self-stabilized lightsails\cite{Lightsail}, and diffractive solar sails\cite{Swartzlander:22, Swakshar:22}.

There are several methods available to calculate optical forces directly from scattering theory, such as the discrete dipole approximation\cite{DDA-force} and the T-matrix method\cite{TMM-force}. However, these methods are designed to calculate scattering from single or simple geometric structure and are generally not well-suited to deal with repeated complex structures or periodic boundary conditions.

The finite-difference time-domain (FDTD), finite-difference frequency-domain (FDFD) and finite element method (FEM) are commonly used for simulating electromagnetic fields for periodic structures by discretising space and adding consideration of boundary conditions. As a time-evolution based method, FDTD can take a long time for a system to reach the steady state, rendering it inefficient for inverse design studies used for force optimization. The FDFD and FEM methods are capable of calculating the steady state directly, but their solution requires the inversion of a huge sparse matrix whose size is determined by the number of lattice sites required. This can be problematic when the structure is 3D and requires a fine mesh to model\cite{FDFD}.

The rigorous coupled-wave analysis (RCWA) method, as a type of Fourier modal method which is optimal for monochromatic situations, has been proven to be an extremely fast and efficient tool for studying reflectance, transmittance and diffraction efficiencies of multilayered periodic structures\cite{RCWA-Formula}. RCWA is not typically used in the field of nanophotonic inverse design due to its slow convergence behaviour in dealing with structures much smaller than a wavelength and high refractive index (RI) contrast materials. Although several approaches based on RCWA\cite{RCWA-DDA, RSTWA, rcwa-nf} have recently demonstrated that these difficulties can be overcome, the suitability of performing optical force calculations using RCWA has, to the best of our knowledge, not been tested.

In this article, we present an optical force simulation approach for periodic structures suitable for nanophotonic inverse design of optical actuators based on the RCWA method. To illustrate our approach, we simulate an experimentally studied structure\cite{benchmark} which acts as a benchmark for comparing force calculations using the RCWA and FDTD methods. After describing the structure studied and outlining the principle of optical force calculation using RCWA and FDTD, we compare optical force calculations of the two methods using two scenarios that are most commonly found in inverse design: changing of incident wavelength and structural parameters. We then investigate the convergence and accuracy for the calculation of optical forces using the two methods. Finally, we demonstrate that the computational time usage of RCWA is a fraction of that of FDTD whilst its hardware requirements are those within the availability of consumer grade laptops.

\section{Benchmark}\label{sec: benchmark} 

\begin{figure}[b]
    \includegraphics[width=.48\textwidth]{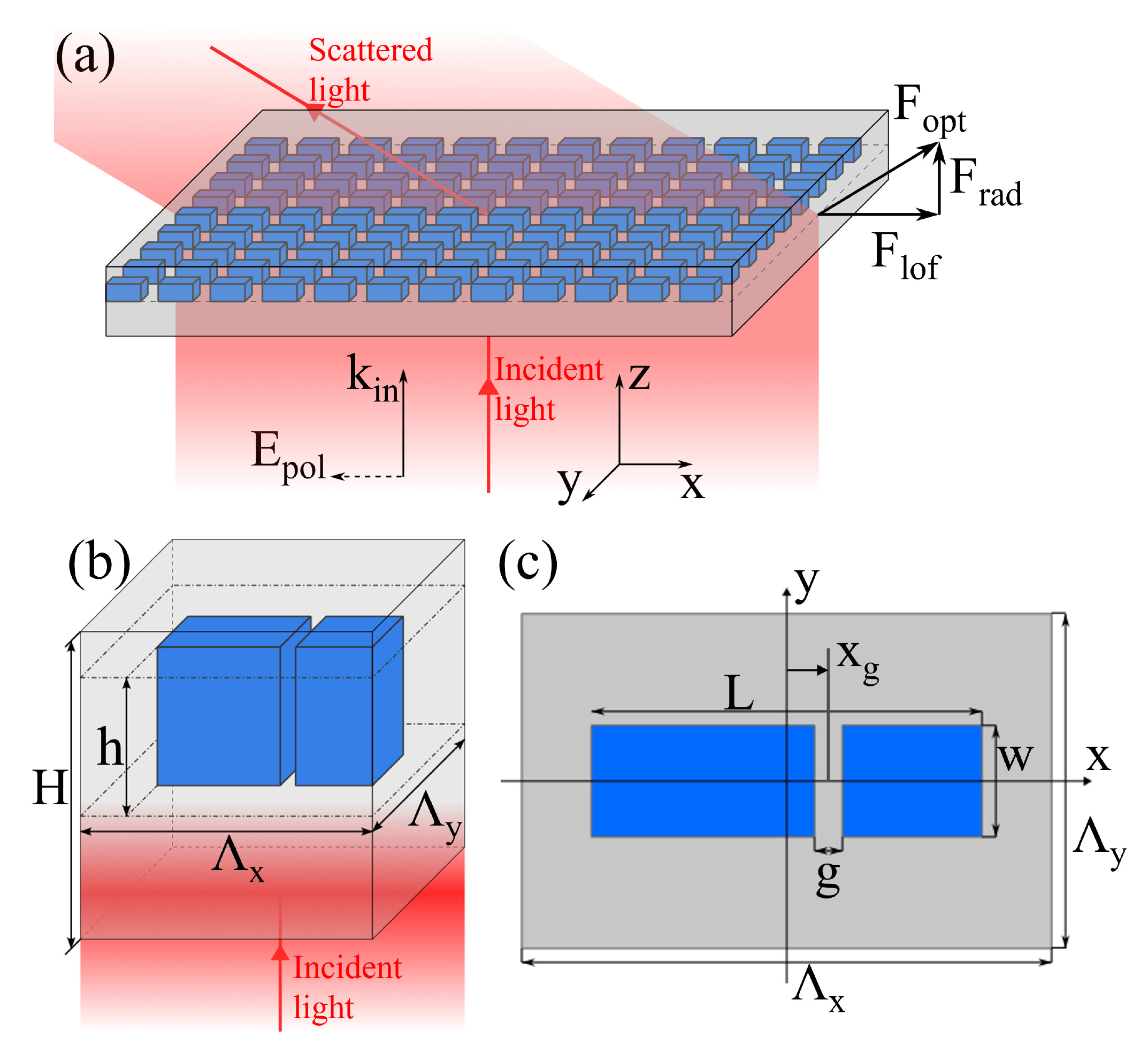}
    \caption{\label{fig: scheme}
        (a) A schematic view of the optical system: linear \emph{x}-polarized light propagating along the \emph{z} direction gets deflected due to the asymmetric dimer inclusions. The resulting optical force $\bi{F}_{\rm opt}$ is decomposed into a lateral optical force $\bi{F}_{\rm lof}$ in $x$ direction and a radiation force $\bi{F}_{\rm rad}$ in $z$ direction. The unit cell structure (b) and its top view (c) demonstrate the definition of coordinates and key structural parameters. For details on dimensions and materials see further details in the main text.
    }
\end{figure}

In order to compare the computation performance of optical forces calculated using RCWA and FDTD, a series of structures derived from the meta-vehicle study of Andr{\'e}n \etal\cite{benchmark} were chosen as a benchmark. These structures consist of an asymmetric repeating motif, as shown in \fref{fig: scheme}, giving rise to a component of optical force perpendicular to the incident field direction.

All the structures were illuminated with a normally incident plane-wave propagating in the $z$ direction and linearly polarized in the $x$ direction as shown schematically in \fref{fig: scheme}(a). Due to the lack of mirror symmetry in the $x$ direction of the system, the light deflected by the structure is asymmetric and as a result exerts an optical force $F_\text{opt}$ with a significant lateral component $F_\text{lof}$ in the $x$ direction and a force component related to the radiation pressure $F_\text{rad}$ in the $z$ direction. As the structures of interest are periodic arrays of nanoantenna inclusions, periodic boundary conditions in the $x$ and $y$ directions of the unit cell are used in the simulation.

\Fref{fig: scheme}(b) and \fref{fig: scheme}(c) show the unit cell structure used in the simulation. The unit cell has a periodicity of $\Lambda_x=\qty{950}{\nm}$ and $\Lambda_y=\qty{600}{\nm}$ and a total thickness of $H=\qty{1000}{\nm}$, and is immersed in water (RI 1.33). The dimer nanoantenna inclusion is \qty{200}{\nm} in width (w), \qty{460}{\nm} in height (h), \qty{720}{\nm} in total length (L) and with a \qty{50}{\nm} gap (g) that separates the two blocks. The inclusion is made of polycrystalline Si (RI 3.45), whereas other parts of the unit cell consist of SiO$_2$ (RI 1.45). The incident wavelength in water $\lambda_w$ and the position of the gap centre $x_g$ were modified and the corresponding optical forces exerted on the structure were studied in the simulations that were performed.

To simplify the simulation, and because the FDTD code, MEEP\cite{MEEP}, currently can only calculate optical forces in a vacuum, we re-scaled all RIs and wavelengths used in this study by the RI of water, such that the background material is reduced to vacuum to allow a direct comparison between the two methods.

\section{Principle of RCWA and FDTD}\label{sec: sim-principle}

Both the FDTD and RCWA methods solve Maxwell's equations, either in the time-domain for FDTD:
\begin{eqnarray}
    \bnabla \times \bi{E}(\bi{r}, t)&= -\partial_t\bi{B}(\bi{r}, t), \label{eq: time-maxwell_e} \\
    \bnabla \times \bi{H}(\bi{r}, t)&= \partial_t\bi{D}(\bi{r}, t)+\bi{J}(\bi{r}, t), \label{eq: time-maxwell_h}
\end{eqnarray}
or in the frequency-domain for RCWA:
\begin{eqnarray}
    \bnabla\times\bi{E}(\bi{r}, \omega)&=-\rmi\omega\bi{B}(\bi{r}, \omega), \label{eq: freq-maxwell_e}\\
    \bnabla\times\bi{H}(\bi{r}, \omega)&= \rmi\omega\bi{D}(\bi{r}, \omega)+\bi{J}(\bi{r}, \omega), \label{eq: freq-maxwell_h}
\end{eqnarray}
where $\bi{r} = (x, y, z)$ is the position vector, $\omega$ is the angular frequency and $\bi{J}$ is the electric current density. The frequency-domain equations are the Fourier transform of the time-domain equations. In both sets of equations, the complexity of solving electric  and magnetic fields $\bi{E}$ and $\bi{H}$ comes from the electric and magnetic flux densities $\bi{D}$ and $\bi{B}$, as they rely on the complex relative permittivity $\tilde{\epsilon}_{\rm r}(\bi{r})$ and permeability $\tilde{\mu}_{\rm r}(\bi{r})$ distribution of the materials:
\begin{eqnarray} 
    \bi{D}(\bi{r})&=\epsilon_0\tilde{\epsilon}_{\rm r}(\bi{r})\bi{E}(\bi{r}), \label{eq: DB_e}\\
    \bi{B}(\bi{r})&=\mu_0\tilde{\mu}_{\rm r}(\bi{r})\bi{H}(\bi{r}), \label{eq: DB_h}
\end{eqnarray}
where $\mu_0$ and $\epsilon_0$ are the permeability and permittivity of vacuum, respectively.

FDTD solves \eref{eq: time-maxwell_e} and \eref{eq: time-maxwell_h} by discretising both space and time and updates the field values at each time step iteratively\cite{FDTD-Intro}. In order to obtain a second-order accuracy for discretization, both time- and space-derivatives are approximated with central finite differences, and different field components are stored on different grid locations known as the Yee Lattice \cite{FDTD-Intro}. The Yee Lattice is designed to allow the $\bi{E}(\bi{r})$ and $\bi{H}(\bi{r})$ components to leap-frog each others in time and space on each iteration. During this process, the electric and magnetic fields are computed as a function of time, so that all the secondary properties such as optical forces and fluxes can be deduced from the primary fields. Since this is a time evolution simulation, it is necessary to include the source current density in the simulation region, and begin from a point in time where there are no fields in space generally.
Consequently, any steady state measurement has to wait for the system to settle, which can be a lengthy process. 

The RCWA method, on the other hand, is a semi-analytical method that solves the frequency-domain equations \eref{eq: freq-maxwell_e} and \eref{eq: freq-maxwell_h}. The current density term can be dropped if the region of interest has no current source in it. By considering a monochromatic incident plane-wave with angular frequency $\omega_0$, or wavenumber $k_0$, and substituting \eref{eq: DB_e} and \eref{eq: DB_h} as well as a reduced magnetic field $\bi{H}_{\rm r}(\bi{r})=-\rmi\sqrt{\mu_0/\epsilon_0}\bi{H}(\bi{r})$, \eref{eq: freq-maxwell_e} and \eref{eq: freq-maxwell_h} are reduced to
\begin{eqnarray} 
    \bnabla\times\bi{E}(\bi{r})&=k_0 \tilde{\mu}_{\rm r}(\bf{r})\bi{H}_{\rm r}(\bi{r}), \label{eq: rcwa-maxwell_e}\\
    \bnabla\times\bi{H}_{\rm r}(\bi{r})&=k_0\tilde{\epsilon}_{\rm r}(\bi{r})\bi{E}(\bi{r}). \label{eq: rcwa-maxwell_h}
\end{eqnarray}
Here, the reduced magnetic field $\bi{H}_{\rm r}(\bi{r})$ is introduced for improved numerical robustness as it shares the same order of magnitude with the electric field $\bi{E}(\bi{r})$.

As a Fourier modal method, RCWA also takes the Fourier transform of space in the $x$ and $y$ directions, which are perpendicular to the beam propagation direction. Hence the fields as well as the permittivity and permeability distribution can be expressed as a Fourier series:
\begin{eqnarray}
     \bi{E}(x,y,z)&=\sum_{m,n}{\bi{e}^{mn}(z)\exp[-\rmi(k_x^{mn}x + k_y^{mn}y)]}, \label{eq: plane wave modes_e}\\
     \bi{H}_{\rm r}(x,y,z)&=\sum_{m,n}{\bi{h}^{mn}(z)\exp[-\rmi(k_x^{mn}x + k_y^{mn}y)]}, \label{eq: plane wave modes_h}
\end{eqnarray}
where $\bi{e}^{mn}(z)$ and $\bi{h}^{mn}(z)$ are $z$-direction analytical amplitude functions of the corresponding Fourier mode. Taking the grating diffraction equation with periodicity $\Lambda_x$ and $\Lambda_y$ into account, the Fourier mode $k_x^{mn}$ and $k_y^{mn}$ are equivalent to the $x$ and $y$ components of the wavevector of the $mn^{\rm th}$ order diffracted plane wave:
\begin{eqnarray} 
    k_x^{mn}&=k_{{\rm in},x} - 2\pi m/\Lambda_x, \\
    k_y^{mn}&=k_{{\rm in},y} - 2\pi n/\Lambda_y,
\end{eqnarray}
where $k_{{\rm in},x}$ and $k_{{\rm in},y}$ are the $x$ and $y$ components of the wavevector of the incident plane wave. Similarly, the relative permittivity and permeability can be expressed in reciprocal space as ${\tilde{\epsilon}_{\rm r}}^{mn}(z)$ and ${\tilde{\mu}_{\rm r}}^{mn}(z)$ such that their distribution only depends on the $z$ coordinate:
\begin{eqnarray}
     {\tilde{\epsilon}_{\rm r}}(x,y,z)&=\sum_{m,n}{\epsilon_{\rm r}^{mn}(z)\exp[-\rmi(k_x^{mn}x + k_y^{mn}y)]}, \\
     {\tilde{\mu}_{\rm r}}(x,y,z)&=\sum_{m,n}{\mu_{\rm r}^{mn}(z)\exp[-\rmi(k_x^{mn}x + k_y^{mn}y)]}.
\end{eqnarray}

Considering a structure can be divided into several layers which are uniform in the $z$ direction, the distribution of $\tilde{\epsilon}_{\rm r}$ and $\tilde{\mu}_{\rm r}$ in real space depends only on the $x$ and $y$ coordinates for each layer. Then, within each layer, after taking the Fourier transform that eliminates the dependence on $x$ and $y$ coordinates, the relative permittivity and permeability in each Fourier mode become constant.

Consequently, the partial derivatives of the $x$ and $y$ components are reduced into algebraic products of a factor of $\rmi k_x^{mn}$ and $\rmi k_y^{mn}$ for the $mn^{\rm th}$ plane wave mode, respectively. Meanwhile, after taking the Fourier transform, the product of the permittivity and the electric field (or the permeability and the magnetic field) becomes a convolution of their Fourier transforms. In numeric computation, the total number of plane wave modes is truncated by $N_{\rm modes}=N_m \times N_n$, where $N_m$ and $N_n$ are the number of modes in the $x$ and $y$ directions, respectively. In doing so the convolution can be replaced by a matrix multiplication by converting the set of constants ${\tilde{\epsilon}_{\rm r}}^{mn}$ and ${\tilde{\mu}_{\rm r}}^{mn}$ into Toeplitz convolution matrices\cite{Numeric} $\lshad\epsilon_{\rm r}\rshad$ and $\lshad\mu_{\rm r}\rshad$; the size of the Toeplitz convolution matrix is $N_{\rm modes} \times N_{\rm modes}$.

Similarly, the $x$ and $y$ components of the wavevector for each plane wave mode $k_x^{mn}$ and $k_y^{mn}$ can be expressed in the matrix forms $\bf{K_x}$ and $\bf{K_y}$. The $x$ and $y$ components of the amplitude functions for each Fourier mode defined in \eref{eq: plane wave modes_e} and \eref{eq: plane wave modes_h} can therefore be written into column vectors $\bi{e}_x(z)$, $\bi{e}_y(z)$ and $\bi{h}_x(z)$, $\bi{h}_y(z)$.

Eventually, \eref{eq: rcwa-maxwell_e} and \eref{eq: rcwa-maxwell_h} reduce to:
\begin{eqnarray} 
    \frac{\rmd}{\rmd z} \left[\begin{array}{c}\bi{e}_x(z) \\ \bi{e}_y(z)\end{array}\right] 
    &=\bf{M}&\left[\begin{array}{c}\bi{h}_x(z) \\ \bi{h}_y(z)\end{array}\right], \label{eq: rcwa-main_e}\\
    \frac{\rmd}{\rmd z} \left[\begin{array}{c}\bi{h}_x(z) \\ \bi{h}_y(z)\end{array}\right] 
    &={\bf N}&\left[\begin{array}{c}\bi{e}_x(z) \\ \bi{e}_y(z)\end{array}\right], \label{eq: rcwa-main_h}
\end{eqnarray}
where $\bf M$ and $\bf N$ are matrices with size $N_{\rm modes} \times N_{\rm modes}$ and defined by:
\begin{eqnarray}
    {\bf M}&=k_0\left[\begin{array}{@{}c@{}c@{}}
       {\bf K_x}\lshad\epsilon_{\rm r}\rshad^{-1}{\bf K_y}&\lshad\mu_{\rm r}\rshad-{\bf K_x}\lshad\epsilon_{\rm r}\rshad^{-1}{\bf K_x} \\
        {\bf K_y}\lshad\epsilon_{\rm r}\rshad^{-1}{\bf K_y}-\lshad\mu_{\rm r}\rshad&-{\bf K_y}\lshad\epsilon_{\rm r}\rshad^{-1}{\bf K_x}
    \end{array}\right]~,\\
    {\bf N}&=k_0\left[\begin{array}{@{}c@{}c@{}}
       {\bf K_x}\lshad\mu_{\rm r}\rshad^{-1}{\bf K_y}&\lshad\epsilon_{\rm r}\rshad-{\bf K_x}\lshad\mu_{\rm r}\rshad^{-1}{\bf K_x} \\
        {\bf K_y}\lshad\mu_{\rm r}\rshad^{-1}{\bf K_y}-\lshad\epsilon_{\rm r}\rshad&-{\bf K_y}\lshad\mu_{\rm r}\rshad^{-1}{\bf K_x}
    \end{array}\right]~.  
\end{eqnarray}

By taking another derivative of \eref{eq: rcwa-main_e} and \eref{eq: rcwa-main_h} with respect to $z$, the association between the electric and magnetic terms is decoupled:
\begin{eqnarray} 
    \frac{\rmd^2}{\rmd z^2} \left[\begin{array}{c}\bi{e}_x(z) \\ \bi{e}_y(z)\end{array}\right]
    -{\bf M}{\bf N}\left[\begin{array}{c}\bi{e}_x(z) \\ \bi{e}_y(z)\end{array}\right]=0, \label{eq: rcwa-2ndode_e}\\
    \frac{\rmd^2}{\rmd z^2} \left[\begin{array}{c}\bi{h}_x(z) \\ \bi{h_y}(z)\end{array}\right] 
    -{\bf N}{\bf M}\left[\begin{array}{c}\bi{h}_x(z) \\ \bi{h}_y(z)\end{array}\right]=0, \label{eq: rcwa-2ndode_h}
\end{eqnarray}
Therefore, within each layer, the two 3D partial differential equations in real space \eref{eq: rcwa-maxwell_e} and \eref{eq: rcwa-maxwell_h} are rewritten into two sets of 1D second order ordinary differential equations with respect to the $z$ coordinate.

In practice, only one set of the equations in \eref{eq: rcwa-2ndode_e} and \eref{eq: rcwa-2ndode_h} need to be solved as the other one can be calculated directly by matrix multiplication; here, we solve for \eref{eq: rcwa-2ndode_e}. Since it is a wave equation in the matrix form, its solution can be found analytically by using eigendecomposition\cite{RCWA-intro}:
\begin{equation}
    \left[\begin{array}{c}\bi{e}_x(z) \\ \bi{e}_y(z)\end{array}\right]
    ={\bf W}e^{-\bkappa z}\bi{c}^+ +{\bf W} e^{+\bkappa z}\bi{c}^-,
\end{equation}
where $\bkappa$ and ${\bf W}$ are the diagonal eigenvalue matrix and the corresponding eigenvector matrix of the matrix ${\bf MN}$. $\bi{c}^+$ and $\bi{c}^-$ represent light that enters into the given layer with forward and backward propagation directions, respectively. The magnetic terms can be calculated from electric terms by using \eref{eq: rcwa-main_h} giving:
\begin{eqnarray}
    \left[\begin{array}{c}\bi{h}_x(z) \\ \bi{h}_y(z)\end{array}\right]
    &=-{\bf NW}\bkappa \rme^{-\bkappa z}\bi{c}^+ + {\bf NW}\bkappa \rme^{+\bkappa z}\bi{c}^- \nonumber \\
    &=-{\bf V} \rme^{-\bkappa z}\bi{c}^+ + {\bf V}\rme^{+\bkappa z}\bi{c}^-,
\end{eqnarray}
where ${\bf V}={\bf NM}\bkappa$. Therefore, the overall solution ${\bPsi}_i(z)$ within the $i^{\rm th}$ layer with thickness $L_i$ can be written as:
\begin{eqnarray}
    {\bPsi}_i(z)&=\left[\begin{array}{@{}c@{}}\bi{e}_{x,i}(z)\\\bi{e}_{y,i}(z)\\\bi{h}_{x,i}(z)\\\bi{h}_{y,i}(z)\end{array}\right]\nonumber\\
    &=\left[\begin{array}{@{}cc@{}} {\bf W}_i&{\bf W}_i\\{\bf -V}_i&{\bf V}_i\end{array}\right]
    \left[\begin{array}{@{}c@{}c@{}}\rme^{-\bkappa_i z}&0\\0&\rme^{\bkappa_i z}\end{array}\right]
    \left[\begin{array}{@{}c@{}}\bi{c}_i^+\\\bi{c}_i^-\end{array}\right].
\end{eqnarray}

The scattering matrix (S-matrix) can be constructed by considering the boundary between the $i^{\rm th}$ and ${(i+1)}^{\rm th}$ layers:
\begin{equation}
    \bPsi_i(z=z_i+L_i)=\bPsi_{i+1}(z=z_{i+1}),
\end{equation}
where $z_i$ and $z_{i+1}$ are the lower boundary coordinates of the corresponding layer. The multiple reflection between each layer can be handled by using the Redheffer star product to stack the S-matrices of each layer to form a global S-matrix that represents the entire structure\cite{RCWA-intro}. Since the solution is expressed in Fourier modes and satisfies the plane-wave decomposition, the diffraction efficiency of each plane wave mode can be found directly by applying the corresponding (reflection and transmission) S-matrix block on the incident light.

\section{Optical force calculation}\label{sec: opt-force}

Optical forces $\bi{F}_{\rm opt}$ typically originate from absorption and scattering of light momentum\cite{ForceIntro}. Direct calculation of optical forces requires knowledge of the electromagnetic nearfield distribution in the vicinity of the structure, which is usually performed by a surface integral of the Maxwell stress tensor $\bsigma$ over a surface $S$ enclosing the object or region of interest:
\begin{equation}
    \bi{F}_{\rm opt} = \oint_S \bsigma \bdot \rmd\bi{A}.
\end{equation}
The components of the Maxwell stress tensor $\sigma_{\alpha\beta}$ in a vacuum can be found easily by using the local electric and magnetic field components:
\begin{equation}
    \sigma_{\alpha\beta} = \epsilon_0 E_\alpha E_\beta + \mu_0 H_\alpha H_\beta - \case{1}{2}\delta_{\alpha\beta}(\epsilon_0 |\bi{E}|^2 + \mu_0 |\bi{H}|^2) ~,
\end{equation}
where $\alpha,\beta \in \{x,y,z\}$. This is the force calculation method used in the FDTD simulation as the nearfield information is easy to obtain for the discretization.

Optical forces $\bi{F}_{\rm opt}$ can also be calculated indirectly by considering the conservation of energy and momentum. The average optical force exerted on a scattering object in one time period of oscillation $T$ can be deduced from the momentum difference of the scattered and incident fields:
\begin{equation}\label{eq: momentum conservation}
    \langle\bi{F}_{\rm opt}\rangle = - \frac{\bi{p}_{\rm scat} - \bi{p}_{\rm in}}{T} ~.
\end{equation}
where $\bi{p}_{\rm scat}$ and $\bi{p}_{\rm in}$ are the momentum of scattered and incident fields, respectively. Since the optical force is linear with the incident power for a given structure, a dimensionless coefficient $\boldsymbol{\mathcal{F}}_{\rm opt}$ is used to characterize the force response resulting from scattering:
\begin{equation}\label{eq: force coefficients}
    \boldsymbol{\mathcal{F}}_{\rm opt} = \frac{\langle\bi{F}_{\rm opt}\rangle}{P_{\rm in}/c},
\end{equation}
where $P_{\rm in}$ is the incident power.

Since the total field calculated by the RCWA method is a summation over all the modes in the plane wave decomposition, the total momentum carried by the scattered field is contributed to linearly by the momentum of each mode. The linear momentum $\bi{p}^{mn}$ of the $mn^{\rm th}$ plane wave mode can be found directly from its unit wave vector $\hat{\bi{k}}^{mn}$ and radiation energy $\mathcal{U}^{mn}$:
\begin{equation}
    \bi{p}^{mn} = \frac{\mathcal{U}^{mn} n_0}{c} \hat{\bi{k}}^{mn} ~,
\end{equation}
where $n_0$ is the RI of the surrounding medium. Therefore, the dimensionless force coefficient $\boldsymbol{\mathcal{F}}_{\rm opt}$ can be found by:
\begin{eqnarray}
    \boldsymbol{\mathcal{F}}_{\rm opt}&=\frac{(\mathcal{U}_{{\rm in}}\hat{\bi{k}}_{\rm in}-\sum_{m,n}{\mathcal{U}^{mn}\hat{\bi{k}}^{mn})n_0/cT}}{P_{\rm in}/c} \nonumber \\
    &=n_0(\hat{\bi{k}}_{\rm in} - \sum_{m,n}{a^{mn} \hat{\bi{k}}^{mn}}),
\end{eqnarray}
where $a^{mn} = P^{mn}/P_{\rm in}$ is the corresponding diffraction efficiency for each mode directly calculated from the RCWA method, $\mathcal{U}_{\rm in}$ is the radiation energy of incident light, and $P^{mn}$ is the power of the $mn^{\rm th}$ plane wave mode. Similar to the optical force $\bi{F}_{\rm opt}$ shown in \fref{fig: scheme}, the dimensionless force coefficient $\boldsymbol{\mathcal{F}}_{\rm opt}$ can also be decomposed into a lateral component $\mathcal{F}_x$ that corresponds to the lateral optical force and a forward component $\mathcal{F}_z$ that corresponds to the radiation pressure.

\section{Implementation}

The FDTD method was performed using MEEP\cite{MEEP}, an open-source software package implemented in C++ with a Python interface.
It was parallelized using the Message Passing Interface (MPI).
The RCWA method was implemented in Python using the PyTorch package for GPU acceleration of matrix operations with CUDA.

Most of simulations were carried out on a single node of BlueCrystal Phase4 (BC4), a high performance computing (HPC) platform at the University of Bristol.
The FDTD simulations on BC4 were performed by two Intel E5-2860 v4 CPU (28 physical cores in total), whereas the RCWA simulations on BC4 were executed by one of the CPUs and an Nvidia P100 GPU with \qty{16}{GB} graphical memory.

Two consumer grade laptops were also used to compare the computing performances of RCWA and FDTD with BC4.
Laptop A has an 8 core Intel i7-11800H CPU and an Nvidia GeForce RTX3080 GPU with \qty{8}{GB} graphical memory, while laptop B has a 12 core Intel i7-1270P CPU and an Nvidia GeForce MX550 GPU with \qty{2}{GB} graphical memory.

\section{Results and Discussion}

\subsection{Force spectrum}\label{sec: variants}
\begin{figure}
    \centering
    \includegraphics[width=.48\textwidth]{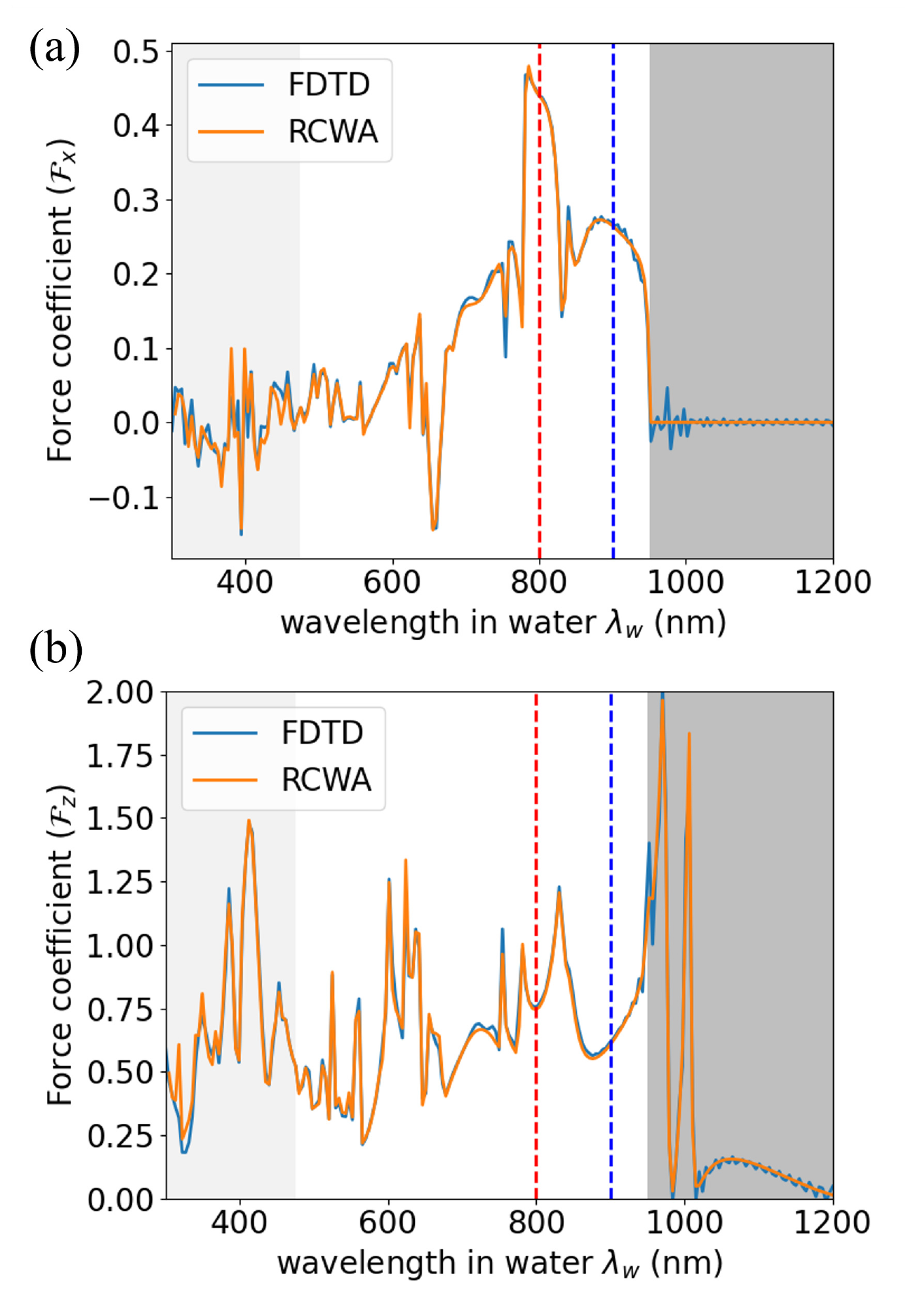}
    \caption{
        Dimensionless force coefficients (a) $\mathcal{F}_x$ and (b) $\mathcal{F}_z$ calculated by FDTD and RCWA for the structure shown in \fref{fig: scheme} with $x_g=\qty{65}{nm}$ and different incident wavelengths. The dark grey background represents the region where only zero-order diffraction exists, whereas the lighter one represents the region where second-order diffraction exists. The red and blue dashed lines mark the two wavelengths in water $\lambda_w=\qty{800}{\nm}$ and $\lambda_w=\qty{900}{\nm}$, respectively, to be investigated in the convergence study .
    }
    \label{fig:spectrum}
\end{figure}

In order to verify that our implementation of the RCWA method provides a correct calculation of the force coefficients,  the force spectrum of the meta-vehicle structure\cite{benchmark} was calculated by varying the incident wavelength using both the RCWA and FDTD approaches. Simulations were performed using the structure shown in \fref{fig: scheme} with the parameters taken from the paper\cite{benchmark} and listed in \sref{sec: benchmark}. \Fref{fig:spectrum}(a) and \fref{fig:spectrum}(b) show the lateral and forward optical force coefficients $\mathcal{F}_x$ and $\mathcal{F}_z$ versus incident wavelength in water, computed for wavelengths ranging from \qty{300}{nm} to \qty{1200}{nm}. Results are superimposed from the FDTD and RCWA approaches showing excellent agreement between the two methods, validating the RCWA implementation.

A number of features of the spectra deserve comment. Due to the periodicity of the structure, and taking the grating diffraction equation into account, in the region where the wavelength in water $\lambda_w>\qty{950}{nm}$, the lateral optical force coefficient $\mathcal{F}_x$ is constantly zero as only the zero order diffraction is permitted and there is no lateral momentum transfer in this region. This region is highlighted by the dark grey background in the right hand region of \fref{fig:spectrum}. It can be seen that the optical forces vary dramatically with the wavelength in water, needing a high sampling density to fully demonstrate details in the spectrum. Any local maxima and minima or narrow peaks and valleys appearing in the spectrum may be an indication of interference effects within the structure or possible resonance effects. For example, the two huge peaks in forward optical force coefficient $\mathcal{F}_z$ spectrum at $\lambda_w=\qty{970.5}{\nm}$ and $\lambda_w=\qty{1006.5}{\nm}$ are due to total (or almost total) reflection. The two valleys located at $\lambda_w=\qty{984.0}{\nm}$ and $\lambda_w=\qty{1191.0}{\nm}$ where $\mathcal{F}_z$ is nearly zero corresponds to total transmission.

The wavelength used by Andr{\'e}n \etal\cite{benchmark} \qty{1064}{\nm} ($\lambda_w=\qty{800}{\nm}$) marked by the red dashed line appears to be a local minimum of $\mathcal{F}_z$ which may result from an interference effect around the inclusion layer. This can be seen in terms of thin film interference by taking the effective refractive index\cite{ERI} (ERI) averaged by volume fraction into account. Since the inclusion layer has the height of $h=\qty{460}{nm}$ and ERI of 2.86, the effective optical path length of the layer is \qty{1316}{nm}, which is about 5/4 of the incident wavelength (\qty{1064}{nm}). This indicates that the light reflected by the inclusion layer has a phase difference of $\pi$ with the incident light. The blue dashed line with $\lambda_w=\qty{900}{\nm}$ marks a place where the optical force coefficient spectrum is rather flat and shows no sign of strong interference or resonance effects. This point will be examined in more detail in the convergence studies (see below).

The force spectrum calculated by FDTD can be performed in a single simulation provided that the light source is broadband and able to cover all wavelengths of interest. This follows from Fourier transforming the time sequence data \cite{FDTD-DFT}. However, one of the disadvantages of this technique is that when the wavelength of interest is too far from the central wavelength of the light source, the fraction of power in that wavelength is small and the error in the force coefficients can be amplified due to machine truncation error. Furthermore, as the time sequence of fields is truncated when the system reaches the criterion for steady state, taking a Fourier transform also produces a termination effect. This is the reason the FDTD curves in \fref{fig:spectrum} show larger noise, especially in the zero-order diffraction region (dark grey region). Although these fluctuations can be reduced by running the simulation for a longer time or applying a window to the time sequence to reduce termination error, the most reliable solution is to run the simulation multiple times for each wavelength individually.

\subsection{Varying gap centre}\label{sec: geom}
\begin{figure}[b]
    \centering
    \includegraphics[width=0.48\textwidth]{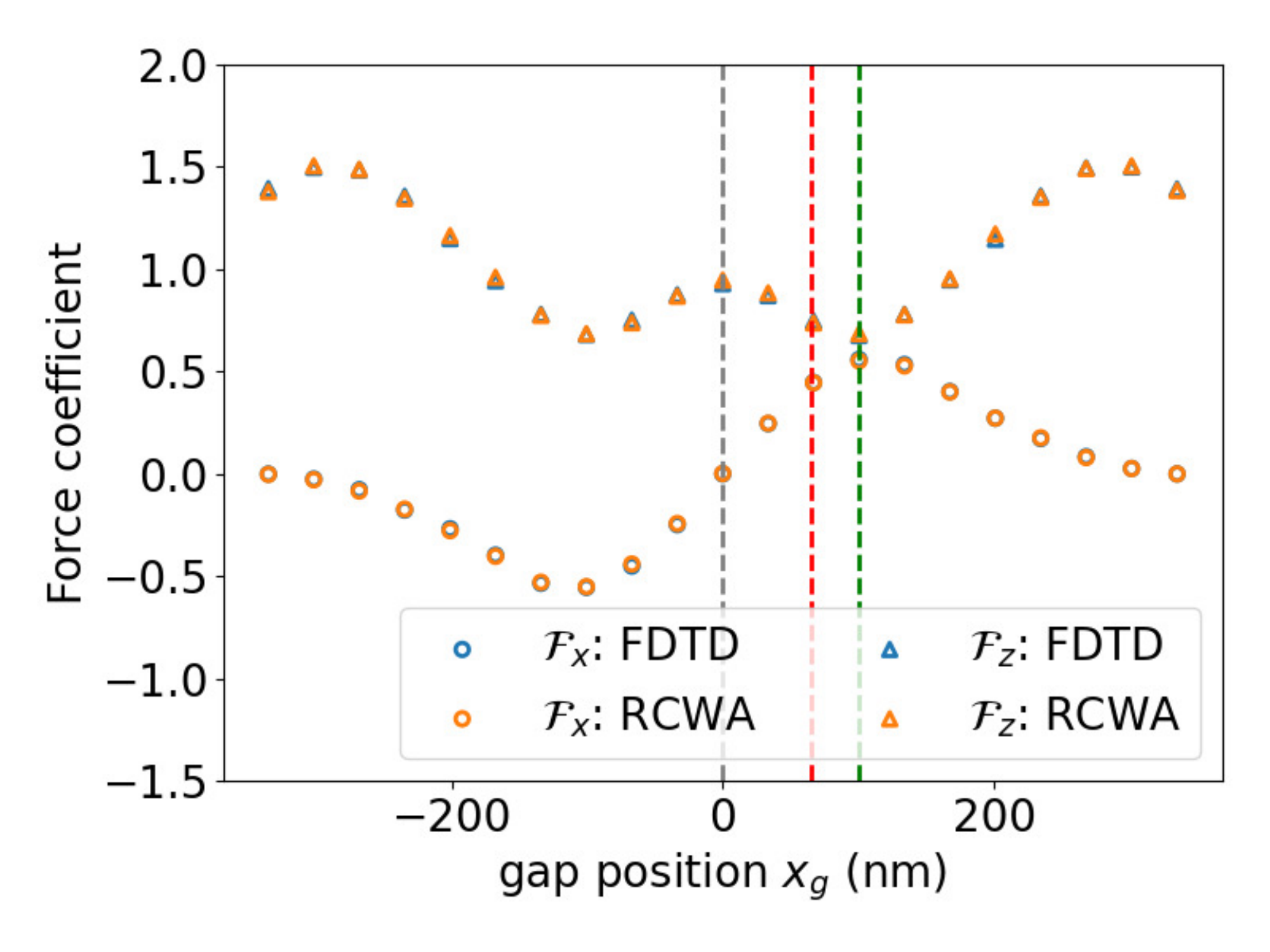}
    \caption{
        Force coefficients $\mathcal{F}_x$ and $\mathcal{F}_z$ calculated by FDTD and RCWA methods with different gap centre position $x_g$ while the total length of the inclusions and the gap width in between are kept unchanged as \qty{720}{nm} and \qty{50}{nm}. The red and green dashed lines indicate gap positions $x_g=\qty{65}{nm}$ and $x_g=\qty{100.5}{nm}$, which correspond to the original structure\cite{benchmark} and a more optimal structure for lateral optical force coefficient $\mathcal{F}_x$ response, respectively.
    }
    \label{fig:geom}
\end{figure}

To further compare the consistency between RCWA and FDTD and their potential behaviour in inverse design, a set of simulations that varies the position of the gap centre $x_g$ (see \fref{fig: scheme}(c)) was performed, while the incident wavelength in water was kept unchanged at $\lambda_w=\qty{800}{nm}$. These simulations, as an example of geometry optimization, were designed to find the optimal gap position for the largest lateral optical force coefficient $\mathcal{F}_x$. \Fref{fig:geom} shows the calculated optical force coefficients versus gap centre position ranging from $x_g=\qty{-335}{nm}$ to $x_g=\qty{335}{nm}$. The total length of the inclusion and the width of the gap in between was kept unchanged as $L=\qty{720}{nm}$ and $g=\qty{50}{nm}$.

The calculated optical force coefficients from RCWA and FDTD show no significant differences. The original structure ($x_g=\qty{65}{nm}$) from Andr{\'e}n \etal\cite{benchmark} is indicated by the red dashed line with $\mathcal{F}_x=0.44$. However, the optimal gap position was found at $x_g=\qty{100.5}{nm}$ as indicated by the green dashed line with $\mathcal{F}_x=0.55$.

In addition, an odd symmetry can be observed for $\mathcal{F}_x$ and an even symmetry for $\mathcal{F}_z$ about the gap position $x_g=\qty{0}{nm}$, which is to be expected due to the fact that, at this gap position, the two blocks on either side of the gap have the same length of \qty{335}{nm} and the mirror symmetry in the $x$ direction is regained. Consequently, the lateral optical force coefficients $\mathcal{F}_x$ is 0 at this point, and for similar reasons at the left and right end as well. 

\subsection{Convergence}\label{sec: convergence}
Since the force spectrum study and the varying gap centre study confirm that the optical force coefficients calculated from the RCWA and FDTD methods are in good agreement, it is important to find the condition for each method where they can work most efficiently. In each case this will be a compromise between the need for the calculated optical force coefficient to be sufficiently accurate whilst the computing resources are kept as small as possible.

The computing resources required and the accuracy of the simulation are controlled directly by the number of lattice sites per \unit{\um} (resolution) for FDTD, and the total number of plane wave modes in each direction (number of modes) for RCWA. Higher resolution or a higher number of modes indicate more precise modelling of the structure, but require greater computing resources in terms of memory and time usage.
\begin{figure}[b]
    \centering
    \includegraphics[width=.48\textwidth]{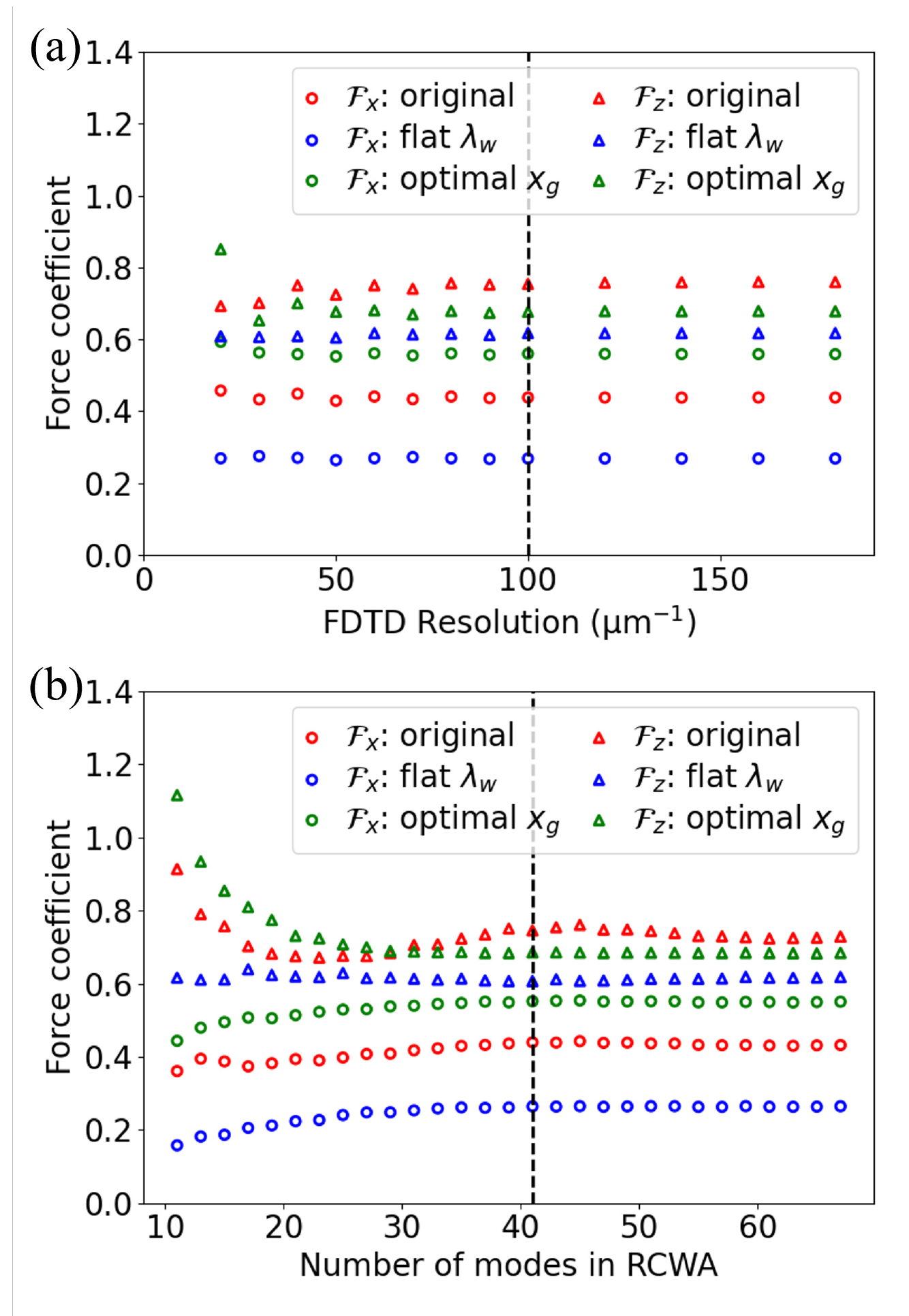}
    \caption{
        Convergence of $x$ and $z$ components of optical force, coefficients $\mathcal{F}_x$ and $\mathcal{F}_z$, using (a) the FDTD method and (b) the RCWA method for three models. The key parameters of the three models are incident wavelength in water $\lambda_w$ and the $x$ coordinate of the gap centre position $x_g$, which can be found in \tref{tab: model}. The black dashed lines represent the condition taken as a standard criterion for convergence and the calculated optical force coefficients at these points are compared in \tref{tab: conv}.
    }
    \label{fig:conv}
\end{figure}

\Fref{fig:conv} shows convergence of the lateral optical force coefficient $\mathcal{F}_x$ and the forward optical force coefficient $\mathcal{F}_z$ of three models as a function of the resolution for FDTD and the number of modes for RCWA. The key parameters of the three models are the incident wavelength in water $\lambda_w$ and the $x$ coordinate of the gap centre position $x_g$, which are summarized in \tref{tab: model}. The ``flat $\lambda_w$" model, as marked by the blue dashed line in \fref{fig:spectrum} with $\lambda_w=\qty{900}{nm}$ and $x_g=\qty{65}{nm}$, represents a region of the force spectrum where possible resonance and interference effects within the structure are less significant and the force response versus wavelength is relatively flat. The ``optimal $x_g$" model, as indicated by the green dashed line in \fref{fig:spectrum} with $\lambda_w=\qty{800}{nm}$ and $x_g=\qty{100.5}{nm}$, corresponds to the optimal gap position with largest lateral optical force coefficient $\mathcal{F}_x$ as discussed previously in \sref{sec: geom}. The ``original" model, as highlighted by the red dashed line in both figures with $\lambda_w=\qty{800}{nm}$ and $x_g=\qty{65}{nm}$, refers to the parameters used by Andr{\'e}n \etal\cite{benchmark}.
\begin{table}[]
\caption{\label{tab: model}
Parameters used in the three convergence studies. The total length of inclusion $L=\qty{720}{nm}$ and the gap width $g=\qty{50}{nm}$ are kept unchanged.
}
    \begin{indented}
        \lineup
        \item[]\begin{tabular}{@{}c@{}cc@{}}
            \br
            Model&Wavelength $\lambda_w$ (\unit{nm})&Gap centre $x_g$ (\unit{nm})\\  
            \mr
            original (red)&800&\065.0\\
            flat $\lambda_w$ (blue)&900&\065.0\\
            optimal $x_g$ (green)&800&100.5\\ 
            \br
        \end{tabular}
    \end{indented}
\end{table}

All the optical force coefficients are varying with a fractional difference between neighbouring data points less than 1\% at and after the \qty{100}{\um^{-1}} resolution for FDTD or 41 modes on each direction ($41^2$ in total) for RCWA, as marked by the black dashed lines in \fref{fig:conv}. Therefore the values at \qty{100}{\um^{-1}} for FDTD and 41 modes for RCWA were taken to represent a standard criterion for convergence across all our simulations, representing a convenient compromise between possible accuracy and available computing resources. The number of modes includes both positive and negative modes and the zero order mode, and hence will always be an odd integer. \Tref{tab: conv} compares the optical force coefficients taken at these standard points and their relative differences ($|\text{RCWA}-\text{FDTD}|/|\text{FDTD}| \times100\%$) as well. It shows that the calculated optical force coefficients from the two methods under standard conditions have good agreement and their differences are all no larger than 2\%.
\begin{table}
    \caption{\label{tab: conv} Converged optical force coefficients computed using the FDTD and RCWA methods, taken at the standard points \qty{100}{\um^{-1}} or 41 modes as shown in \fref{fig:conv}. The differences between the two methods in the three cases are also shown and are all within 2\%.}
    \begin{indented}
        \lineup
        \item[]\begin{tabular}{cccc}
            \br
            Model  & FDTD    & RCWA    & Differences \\
            \mr
            $\mathcal{F}_x$: original&0.4378&0.4396& 0.41\%         \\
            $\mathcal{F}_z$: original&0.7536&0.7454& 1.09\%         \\ 
            $\mathcal{F}_x$: flat $\lambda_w$&0.2679&0.2643&1.33\%         \\
            $\mathcal{F}_z$: flat $\lambda_w$&0.6169&0.6069&1.62\%         \\ 
            $\mathcal{F}_x$: optimal $x_g$&0.5596&0.5519&1.39\%         \\
            $\mathcal{F}_z$: optimal $x_g$&0.6761&0.6852&1.36\%         \\ 
            \br
        \end{tabular}  
    \end{indented}
\end{table}

Close examination shows that the fluctuation of the RCWA results for the $\mathcal{F}_z$ curve of the ``original" model after the chosen convergence point is slightly larger than for the FDTD method. This variation cannot be removed by simply adding more modes and it appears to be due to the Gibbs phenomenon which occurs when taking Fourier transforms of discontinuous systems. In the present case we are dealing with a discontinuous permittivity and permeability at the boundary of different materials in our test structure. This is the main difficulty for RCWA when dealing with structures with characteristic sizes which are much smaller than the incident wavelength. Resonance or interference effects in the system at certain wavelengths might also amplify such fluctuations, which could explain the $\mathcal{F}_z$ curve of the ``original" model in the RCWA method in \fref{fig:conv}.

Convergence of the near field in RCWA has previously been discussed and shown to be much more problematic than for the far field\cite{rcwa-nf}. This is because the near field region relies on accurate calculation of both propagating diffraction orders and evanescent modes, which are particularly susceptible to the Gibbs phenomenon\cite{rcwa-nf}. Thus the suitability of direct optical force calculation using the near field results from RCWA using, for example, the Maxwell stress tensor, is questionable. However, the influence of the Gibbs phenomenon on far field properties such as diffraction efficiencies can be controlled within a reasonable degree as seen in \fref{fig:conv}, and the consequent fluctuations in the optical forces calculated indirectly are within 1\%. Therefore, we conclude that RCWA is suitable for indirect optical force calculation using momentum conservation, as in \eref{eq: momentum conservation} and \eref{eq: force coefficients}.

Concerning the resolution limit for FDTD, any lower resolution than \qty{100}{\um^{-1}} is not a reasonable choice for the converged point because the \qty{100}{\um^{-1}} resolution only gives 5 lattice sites to represent the \qty{50}{nm} gap. Thus any lower resolution cannot describe the geometry of the structure adequately.

\subsection{Speed and computing resources}
In order to test the suitability of the RCWA method for optical force optimization, aside from the ability of yielding accurate results, the speed and computing resource requirement must also be considered. During the simulations mentioned above, we observed that the RCWA method is typically on the order of 10 times faster than the FDTD method on BC4 for a given force calculation, whilst its computing resource requirements are generally within the availability of a consumer grade laptop.

\begin{figure}[b]
    \centering
    \includegraphics[width=.46\textwidth]{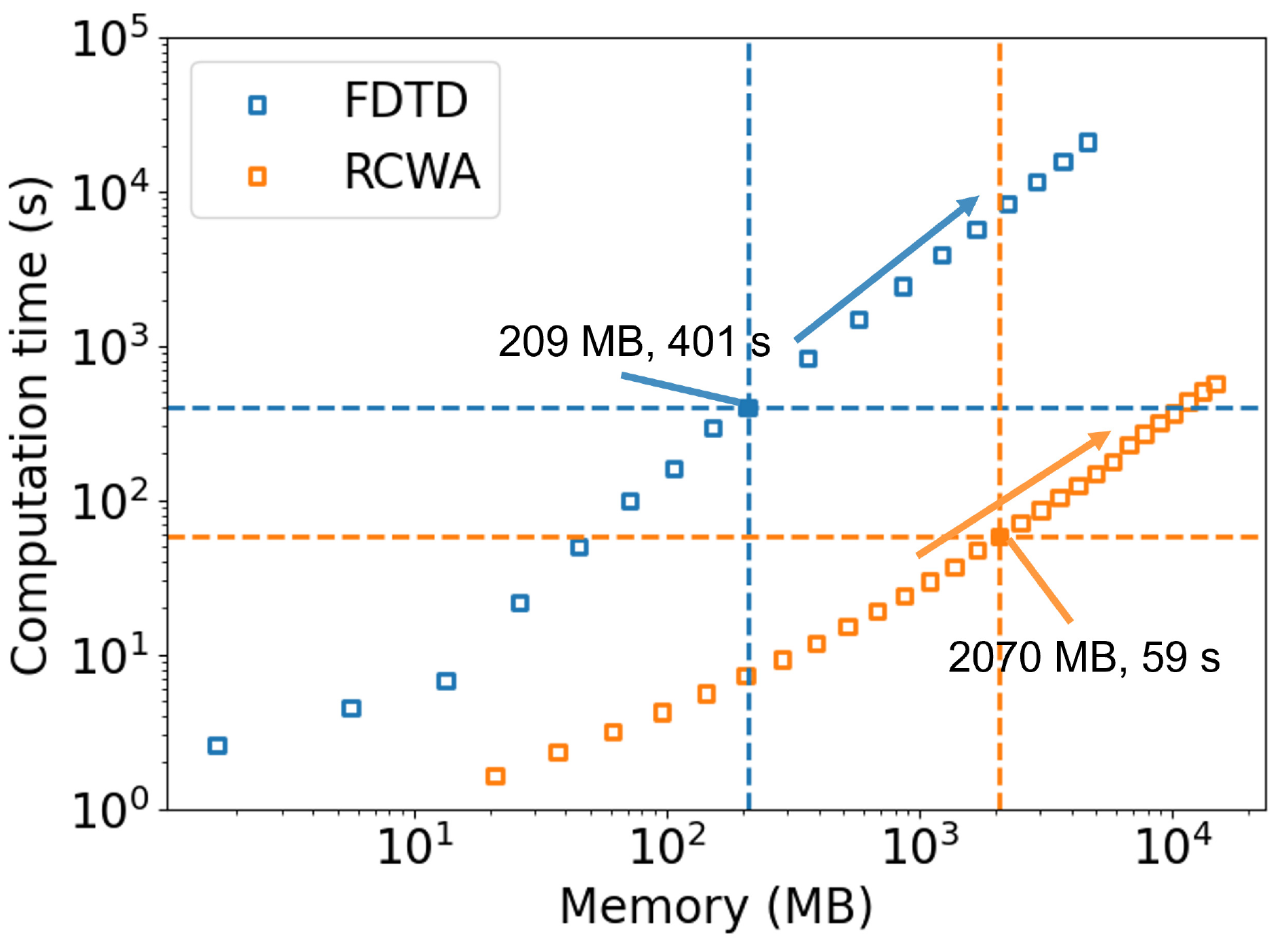}
    \caption{\label{fig:time}
    Computational time usage (seconds) versus memory usage (MB) for the ``original" model with parameters shown in \tref{tab: model} using the FDTD method with different resolution and the RCWA method with different number of modes performed on BC4. The solid points represent the time and memory usage taken at the standard points \qty{100}{\um^{-1}} and 41 modes as shown in \fref{fig:conv}. Arrows indicate increasing resolution or number of modes.
    }
\end{figure}

\Fref{fig:time} shows the time usage on BC4 for the ``original" model in the convergence study with respect to memory usage. It shows that the time usage of RCWA method for a single force evaluation is a fraction of that of the FDTD method. Simulations of other cases perform similarly, \emph{i.e.}, the time usage at the standard points, \qty{100}{\um^{-1}} resolution for FDTD and 41 modes for RCWA, take roughly 400 seconds and 60 seconds on BC4, respectively.

In the force spectrum study, the time usage of the FDTD simulation with a broadband source is about 35 minutes on BC4, compared with roughly 201 minutes (3.5 hours) for the RCWA method. RCWA is much slower than the FDTD in this situation because it cannot take advantage of Fourier transforming the time sequence data, as explained previously, and therefore it has to simulate for each wavelength of interest individually. The 201 minutes time usage is expected because there were 201 different wavelengths involved in \fref{fig:spectrum}, and each RCWA simulation takes about 1 minute. On the other hand, the FDTD time usage in \fref{fig:spectrum} is greater than its typical value, \emph{i.e.} 400 seconds, because the broadband light source needs longer for the simulation to reach a steady state. However, a disadvantage of Fourier transforming the time series data from the FDTD simulation is that it introduces termination effects in the force spectrum, and the simplest way to eliminate them is to run the FDTD simulation for each wavelength individually, which would take about 28 hours on BC4.

In the gap centre variation study (\fref{fig:geom}), for both RCWA and FDTD, each geometry was investigated with a single separate simulation. As there were 21 different geometries in the study, the time usage for the FDTD method was about 3.25 hours whilst the RCWA method only took about 21 minutes. Therefore the average time usage for a single simulation in this case is about 550 seconds for FDTD and 60 seconds for RCWA. The slowing of the FDTD method compared with the standard case, appears to be because some geometries take longer to reach the criterion of steady state than others. Overall, these timings indicate that for a typical inverse design problem requiring force responses for a large number of different structures, the RCWA method is around an order of magnitude more efficient than the FDTD method.

Given the apparent speed benefit of RCWA when running on a HPC platform, we decided to test the feasibility of running the same code on more accessible, consumer grade, equipment. \Tref{tab:compare} shows a time usage comparison of two laptops with BC4 for our ``original" model at the standard point (\qty{100}{\um^{-1}} resolution for FDTD and $41\times41$ modes for RCWA). Single precision (64bits) and double precision (128bits) data types were tested because the graphics cards on BC4 are optimized for double precision whereas the graphics cards of the most commonly available consumer grade laptops are configured for predominantly single precision arithmetic.
\begin{table*}
    \caption{\label{tab:compare}
    Time usage comparison of FDTD and RCWA method on the original model ($\lambda_w=\qty{800}{\nm}$, $x_g=\qty{65}{nm}$) with \qty{100}{\um^{-1}} resolution (FDTD) and $41\times41$ modes (RCWA) on different machines.
    }
    \begin{indented}
    \lineup
        \item[]\begin{tabular}{@{}cccc@{}}
            \br
            Brand & BlueCrystal Phase4 (HPC) & Laptop A  & Laptop B \\ 
            CPU: cores&Intel E5-2680 v4: 28 cores&Intel i7-11800H: 8 cores&Intel i7-1270P: 12 cores\\
            GPU: memory&Nvidia Tesla P100: 16 GB&Nvidia GeForce RTX3080: 8 GB&Nvidia GeForce MX550: 2GB\\
            \mr
            Time FDTD (seconds)&\0401.0&1701.8&1563.0\\
            Time RCWA-single (seconds)&\0\050.4&\0\049.6&\0\054.6\\
            Time RCWA-double (seconds)&\0\058.6&\0\092.2&---\\
            \br
        \end{tabular}
    \end{indented}
\end{table*}

The single precision RCWA results show that the time usage on both laptops is about the same as BC4 and could be even faster if its GPU had better performance. Although the double precision RCWA method on Laptop A is 1.5 times slower, it is still about 20 times faster than the FDTD method on the laptop and more than 4 times faster than the FDTD method on BC4. Therefore, it is clear that the RCWA method can also work on consumer grade computers and laptops with a good performance. This strengthens the suitability of RCWA for optical force optimization, because access to HPC systems tends to be restricted or expensive, while CUDA capable graphics cards are readily available.

Single precision variables are not recommended for use in practice, as they bring higher truncation and rounding error which would potentially be amplified by the interference effects discussed above. \Fref{fig:time} also indicates that the RCWA method is more memory-intensive than the FDTD method, and its memory usage is mainly the graphical memory. The missing data in \tref{tab:compare} is due to the fact that running 41 modes in double precision requires \qty{2070}{MB} graphical memory which is larger than the graphical memory capacity (\qty{2}{GB}) of Laptop B. The available graphical memory of a GPU limits the largest number of modes in each direction that can be used on it, and it is important to estimate the maximum graphical memory usage during the computation. The graphical memory usage (in bytes) is $768N^4$ for double precision and $384N^4$ for single precision for a given number of modes, $N$, on each direction. This suggests an upper limit for double precision work of $N=39$ for the \qty{2}{GB} GPU in Laptop B and $N=57$ for the \qty{8}{GB} GPU in Laptop A.

Therefore, the RCWA method is more suitable for fast optical force simulation than FDTD on a consumer grade laptop with a CUDA capable graphics card and a graphical memory larger than \qty{2}{GB}, which opens up opportunities for optical force based inverse design and geometry optimization without the need of HPC resources.

\section{Conclusion}
This paper shows that optical force coefficients can be calculated using the RCWA method giving results which are consistent with the FDTD method for a fraction of the computer time.
The FDTD method converged when the resolution reached \qty{100}{\um^{-1}}, with a calculation time about 400 seconds on the local HPC system.
On the other hand, the RCWA method converged when the number of modes in each direction reached 41, and took about 60 seconds on the same HPC system.
The differences in optical force coefficients for the benchmark case between RCWA and FDTD methods are no larger than 2\%.

The RCWA method is capable of running on both HPC and consumer grade laptops within a reasonable time as long as the laptop has a CUDA-compatible GPU and graphical memory larger than 2GB (for double precision and 41 modes on each direction).
RCWA is typically about 10 times faster than FDTD on HPC system such as BC4, and about 20 times faster on consumer grade laptops.
The FDTD method could be made faster by using more HPC nodes, but such facilities are not generally available.
Therefore, the RCWA method is very suitable for studies that include calculating optical force responses for a large number of distinct structures.
Furthermore, in the consideration of green computing, using the RCWA method on personal laptops could save considerable resources and energy compared to using the HPC, and is therefore the more sustainable option.

\ack
BG acknowledges the China Scholarship Council (CSC) and University of Bristol for a postgraduate scholarship (award number 202008060217),
and Advanced Computing Research Centre (ACRC) at the University of Bristol for High Performance Computing platform BlueCrystal phase4 (BC4) support.
HG acknowledges support through an Impact Acceleration Account co-funded by EPSRC and Bristol Nano Dynamics (grant number EP/R511663/1).

\section*{References}
\bibliography{reference.bib}

\end{document}